\documentstyle[seceq,preprint]{ptptex}

\newcommand{\A}{\mbox{\boldmath$A$}}
\newcommand{\n}{\mbox{\boldmath$n$}}
\newcommand{\ms}{\hspace{-1mm}}

\title{Consistency Conditions of the Faddeev-Niemi-Periwal Ansatz\\
for the {\boldmath $SU(N)$} Gauge Field}

\author{Minoru {\sc Hirayama}, Mika {\sc Kanno},
Masataka {\sc Ueno}\\
and\\
 Hitoshi {\sc Yamakoshi}$^*$}

\inst{Department of Physics, Toyama University, Toyama 930-8555, Japan
\\
$^*$Toyama National College of Technology, Toyama 939-8630, Japan}

\notypesetlogo

\recdate{December 19, 1998}

\abst{The consistency condition of the Faddeev-Niemi ansatz for the 
gauge-fixed massless $SU(2)$ gauge field is discussed.  
The generality of the ansatz is demonstrated by obtaining a sufficient condition for
the existence of the three-component field introduced by Faddeev and Niemi.
It is also shown that the consistency conditions determine this
three-component field as a functional of two arbitrary functions.
The consistency conditions corresponding to the Periwal ansatz for the 
$SU(N)$ gauge field with $N\geq3$ are also obtained.
It is shown that the gauge field obeying the Periwal ansatz must satisfy 
extra $(N-1)(N-2)/2$ conditions.}

\begin{document}

\maketitle

\section{Introduction}
Recently, Faddeev and Niemi proposed an interesting parametrization for a
four dimensional $SU(2)$ gauge field.\cite{rf:1} \
They assert that a gauge-fixed massless $SU(2)$ gauge field 
$\A_\mu(x)=(A^1_\mu(x),A^2_\mu(x),A^3_\mu(x))$ can be expressed as
\begin{equation}
\A_\mu(x)=C_\mu(x)\n(x)+\rho(x)\partial_\mu\n(x)
+\{1+\sigma(x)\}\partial_\mu\n(x)\times\n(x),
\end{equation} 
where $\n(x)=(n^1(x),n^2(x),n^3(x))$, $C_\mu(x)$, $\rho(x)$ and $\sigma(x)$
are a three-component scalar field satisfying
\begin{equation}
\n(x)^2=n^a(x)n^a(x)=1,
\end{equation} 
a massless gauge-fixed vector field, and two scalar fields, respectively.
$\A_\mu(x)$ has six physical degrees of freedom. This is equal to the sum of
those of $\n(x)$, $C_\mu(x)$, $\rho(x)$ and $\sigma(x)$: 6=2+2+1+1. 
Faddeev and Niemi stress that the field $\A_\mu(x)$ is convenient to describe the 
ultraviolet limit, while the fields $\n(x)$, $C_\mu(x)$, $\rho(x)$ and 
$\sigma(x)$ are appropriate for description of infrared limit. 
It should be mentioned that the classical action for $\n(x)$ induced from the 
conventional gauge invariant action for $\A_\mu(x)$ gives rise to some 
interesting knot-like soliton configurations,\cite{rf:2} realising Kelvin's 
idea of a vortex atom.\cite{rf:3} 
Since the apparent degrees of freedom of the left- and the right-hand sides of 
(1$\cdot$1) are different, there must be some consistency conditions.
If these consistency conditions are satisfied for any gauge
configuration, the ansatz (1$\cdot$1) is complete as Faddeev and Niemi
asserted. On the other hand, if the consistency conditions give rise to some
restrictions on $\A_\mu(x)$, the expression (1$\cdot$1) is of limited use.
We investigate here the consistency condition from a viewpoint different
from that of Faddeev and Niemi. Another problem is to find a method to
obtain the field $\n(x)$ from given $\A_\mu(x)$.
This problem is not touched upon in Ref. 1).
We thus consider two questions in this paper: (1) what are the consistency
conditions implied by (1$\cdot$1)? 
(2) How is the field $\n(x)$ determined by $\A_\mu(x)$?  
It turns out that the existence conditions for $\rho(x)$ and $\sigma(x)$ play
the role of the consistency conditions.
These conditions can be rewritten as a system of differential equations for 
$\n(x)$. The system of differential equations, in principle, determine
$\n(x)$ from $\A_\mu(x)$ if its integrability condition is satisfied.
Investigating the integrability of this system, we find a sufficient
condition for the field $\n(x)$ to exist.
This sufficient condition is rather loose, meaning that the ansatz can be
used for a wide class of gauge configurations. Furthermore, noting that
the ansatz might be applicable even if the above sufficient condition is
not satisfied, we are led to the expectation that the Faddeev-Niemi Ansatz
can be used generally.

On the other hand, the extension of the Faddeev-Niemi Ansatz to the case of  
the gauge group $SU(N)$ with $N\geq3$ was discussed by Periwal.\cite{rf:4}
Although another ansatz was proposed by Faddeev and Niemi quite recently,
\cite{rf:5} we here investigate Periwal's proposition.  
He proposed a parametrization for the gauge-fixed massless $SU(N)$ gauge field 
which seems to lead to a matching of the degree of freedom for any $N$.
In Periwal's ansatz, the field $\n(x)$ in the case of $SU(2)$ is replaced by
$N-1$ commuting traceless Hermitian $N\times N$ matrices $H_I(x)$ 
$(I=1,2,\cdots,N-1)$ which possess $N^2-N+(N-1)(N-2)/2$ degrees of freedom. 
He also introduces $N-1$ massless gauge-fixed Abelian gauge fields 
$C_{\mu,I}(x)$ possessing $2(N-1)$ degrees of freedom, $N-1$ scalar fields 
$\phi_I(x)$, and $N(N-1)/2$ scalar fields $\beta_{IJ}(x)=\beta_{JI}(x)$ 
$(I,J=1,2,\cdots,N-1)$. 
The total number of degrees of freedom possessed by $H_I(x)$, $C_{\mu,I}(x)$,
$\phi_I(x)$ and $\beta_{IJ}(x)$ is $2(N^2-1)$. This is equal to the number of 
degrees of freedom possessed by the massless gauge-fixed $SU(N)$ gauge field.
Thus, if all of these fields are in fact independent, his ansatz realizes a 
correct matching of degrees of freedom.
In this paper, we seek the consistency conditions for the Periwal ansatz 
and find $3(N-1)(N+2)/2$ conditions. 
It is also pointed out that the $SU(N)$ gauge field obeying Periwal's ansatz
satisfies extra $(N-1)(N-2)/2$ conditions due to the $SO(N-1)$ symmetry
hidden in his ansatz.

This paper is organized as follows. In \S2, we consider the case of $SU(2)$.
We find some expressions of the consistency conditions in terms of $\n(x)$
and $\A_\mu(x)$. 
Except for very special cases, we find that these conditions can fix the 
field $\n(x)$ and conclude that the Faddeev-Niemi ansatz is applicable.
In \S3, the case of $SU(N)$ with $N\geq3$ is considered. 
The final section, \S4, is devoted to summary and discussions.
Two appendices are included to describe some details of the calculation.

\section{The case of $SU(2)$}

As mentioned in \S1, the ansatz (1$\cdot$1) should be associated with 
some consistency conditions, which we seek in this section.
We show that these conditions yield a system of differential equations 
from which the field $\n(x)$ is determined. 

\subsection{Consistency conditions implied by the Faddeev-Niemi Ansatz}

Originally, the field $\A_\mu(x)$ has twelve components,
while it has only six physical degrees of freedom,
because it describes a gauge-fixed massless iso-vector field. 
In \S1, we mentioned that the number of degrees of freedom carried by $\A_\mu(x)$ 
is equal to the sum of those of the new variables $\n(x)$, $C_{\mu}(x)$, 
$\rho(x)$ and $\sigma(x)$.  

Since the three-component field $\n(x)$ is assumed to satisfy (1$\cdot$2), 
we have 
\begin{equation}
\n\cdot\partial_\mu\n=0, \ \n\cdot(\partial_\mu\n\times\n)=0, \ 
\partial_\mu\n\cdot(\partial_\mu\n\times\n)=0.
\end{equation}
Equations (1$\cdot$1) and (2$\cdot$1) lead to the relations
\begin{subeqnarray}
C_\mu \hspace{-3mm}& = &\hspace{-3mm} \n\cdot\A_\mu,\\
\rho \hspace{-3mm}& = & \hspace{-3mm}
\frac{\partial_\mu\n\cdot\A_\mu}{(\partial_\mu\n)^2},\\
1+\sigma \hspace{-3mm}& = &\hspace{-3mm}
\frac{(\partial_\mu\n\times\n)\cdot\A_\mu}{(\partial_\mu\n\times\n )^2}, 
\end{subeqnarray}
where $\mu$ should be set equal to 0, 1, 2 or 3, and the sum over 
$\mu$ = 0, 1, 2, 3 should not be taken on the right-hand sides of (2$\cdot$2b) and
(2$\cdot$2c).
For the fields $\rho(x)$ and $\sigma(x)$ to exist, the fields $\A_\mu(x)$ and 
$\n(x)$ must satisfy the following conditions:
\begin{equation}
\frac{\partial_{0}\n\cdot\A_{0}}{(\partial_{0}\n)^2}=
\frac{\partial_{1}\n\cdot\A_{1}}{(\partial_{1}\n)^2}=
\frac{\partial_{2}\n\cdot\A_{2}}{(\partial_{2}\n)^2}=
\frac{\partial_{3}\n\cdot\A_{3}}{(\partial_{3}\n)^2},
\end{equation}
\begin{equation}
\frac{(\partial_{0}\n\times\n)\cdot\A_{0}}{(\partial_{0}\n\times\n )^2}=
\frac{(\partial_{1}\n\times\n)\cdot\A_{1}}{(\partial_{1}\n\times\n )^2}=
\frac{(\partial_{2}\n\times\n)\cdot\A_{2}}{(\partial_{2}\n\times\n )^2}=
\frac{(\partial_{3}\n\times\n)\cdot\A_{3}}{(\partial_{3}\n\times\n )^2}.
\end{equation}
The six relations in (2$\cdot$3) and (2$\cdot$4) are the consistency 
conditions associated with the ansatz (1$\cdot$1).
It is not clear at the present stage of discussion, however, how the field 
$\n(x)$ is obtained from $\A_\mu(x)$.
 
If we introduce the variables $\theta(x)$ and $\varphi(x)$ defined by 
$\n(x)=(\sin\theta(x)\cos\varphi(x), \\ \sin\theta(x)\sin\varphi(x), \ 
\cos\theta(x))$, \ Eqs. (2$\cdot$3) and (2$\cdot$4) become
\begin{equation}
M_\mu\left(\begin{array}{c}
G_\nu\\
H_\nu\\
\end{array}\right) 
=M_\nu\left(\begin{array}{c}
G_\mu\\
H_\mu\\
\end{array}\right), 
\end{equation}
where $M_{\mu}(x)$, $G_{\mu}(x)$ and $H_{\mu}(x)$ are given by
\begin{equation}
M_\mu =\left(\begin{array}{cc}
\partial_\mu\theta & -\sin\theta\hspace{1mm}\partial_\mu\varphi\\
\sin\theta\hspace{1mm}\partial_\mu\varphi & \partial_\mu\theta
\end{array}\right), 
\end{equation}
\begin{equation}
G_\mu=\frac{\partial\n}{\partial\theta}\cdot\A_\mu,\hspace{3mm}
H_\mu=\frac{1}{\sin\theta}\frac{\partial\n}{\partial\varphi}\cdot\A_\mu.
\end{equation}
Since $M_{\mu}(x)$ satisfies $M_{\mu}(x) M_{\nu}(x)=M_{\nu}(x) M_{\mu}(x)$, 
only three equations in 
(2$\cdot$5), with, e.g. $\mu=0$, $\nu=1,\ 2,\ 3$, are independent.
With the help of the relations 
\begin{subeqnarray}
\hspace{-25mm}& \left(\begin{array}{cc}
a & -b\\
b & a
\end{array}\right) &\hspace{-3.5mm}
\left(\begin{array}{c}
u\\
v
\end{array}\right)
=\left(\begin{array}{cc}
u & -v\\
v & u
\end{array}\right)
\left(\begin{array}{c}
a\\
b
\end{array}\right),\hspace{3mm}a,\ b,\ u,\ v\in\mbox{\boldmath$C$},\\
\hspace{-25mm}\nonumber\\
\hspace{-25mm}& \left(\begin{array}{cc}
a & -b\\
b & a
\end{array}\right) &\hspace{-3.5mm}
\left(\begin{array}{cc}
a' & -b'\\
b' & a'
\end{array}\right)
=\left(\begin{array}{cc}
a' & -b'\\
b' & a'
\end{array}\right)
\left(\begin{array}{cc}
a & -b\\
b & a
\end{array}\right),\hspace{3mm}a,\ b,\ a'\ ,b'\in\mbox{\boldmath$C$},
\end{subeqnarray}
we obtain
\begin{equation}
\left(\begin{array}{cc}
G_\mu & -H_\mu\\
H_\mu &G_\mu
\end{array}\right)^{-1}
\left(\begin{array}{c}
\partial_\mu\theta\\
\sin\theta\hspace{1mm}\partial_\mu\varphi
\end{array}\right)
=\left(\begin{array}{cc}
G_\nu & -H_\nu\\
H_\nu &G_\nu
\end{array}\right)^{-1}
\left(\begin{array}{c}
\partial_\nu\theta\\
\sin\theta\hspace{1mm}\partial_\nu\varphi
\end{array}\right)
\end{equation}
from (2$\cdot$5). We find that the l.h.s. (r.h.s.) of (2$\cdot$9)
sould be independent of $\mu$ $(\nu)$. Then we have
\begin{equation}
\left(\begin{array}{c}
\partial_\mu\theta\\
\sin\theta\hspace{1mm}\partial_\mu\varphi
\end{array}\right)
=\left(\begin{array}{cc}
G_\mu & -H_\mu\\
H_\mu &G_\mu
\end{array}\right)
\left(\begin{array}{c}
\beta\\
-\gamma
\end{array}\right),
\end{equation} 
where the functions $\beta(x)$ and $\gamma(x)$ are independent of $\mu$.
We see that (2$\cdot$8a) and (2$\cdot$10) yield
\begin{equation}
\left(\begin{array}{c}
\partial_\mu\theta\\
\sin\theta\hspace{1mm}\partial_\mu\varphi
\end{array}\right)
=L\left(\begin{array}{c}
G_\mu\\
H_\mu
\end{array}\right),\hspace{3mm}
L=\left(\begin{array}{cc}
\beta & \gamma\\
-\gamma & \beta
\end{array}\right),
\end{equation}
where the functions $\beta(x)$ and $\gamma(x)$ should be regarded as arbitrary
functions. It can be readily checked that $\partial_\mu\theta(x)$ and
$\partial_\mu\varphi(x)$ given by (2$\cdot$11) with arbitrary $\beta(x)$ and
$\gamma(x)$ satisfy (2$\cdot$5). We note that the $\mu$-independent
functions $\beta(x)$ and $\gamma(x)$ are related to $\A_\mu(x)$ and $\n(x)$ by
\begin{subeqnarray}
\beta\hspace{-3.5mm} & = & \hspace{-3.5mm}
\frac{G_\mu\partial_\mu\theta+H_\mu\sin\theta\hspace{1mm}\partial_\mu\varphi}
{G_{\mu}^{2}+H_{\mu}^{2}},\\
\gamma\hspace{-3.5mm}& = &\hspace{-3.5mm} 
\frac{H_\mu\partial_\mu\theta-G_\mu\sin\theta\hspace{1mm}\partial_\mu\varphi}
{G_{\mu}^{2}+H_{\mu}^{2}}.
\end{subeqnarray}

\subsection{Integrability condition for Eq. (2$\cdot$11)}
We next consider the integrability condition of (2$\cdot9$) which insures 
the existence of $\theta(x)$ and $\varphi(x)$. The requirements
\begin{equation}
\partial_\mu(\partial_\nu\theta)=\partial_\nu(\partial_\mu\theta),\hspace{3mm}
\partial_\mu(\partial_\nu\varphi)=\partial_\nu(\partial_\mu\varphi),
\end{equation}
lead to the integrability condition
\begin{equation}
\left\{\begin{array}{c}
\partial_\mu(LA_{\nu}^{a})-\partial_\nu(LA_{\mu}^{a}) 
+\epsilon^{abc}A_{\mu}^{b}A_{\nu}^{c}L^{2}
\left(\begin{array}{cc}
0 & 1\\
-1 & 0
\end{array}\right)
\end{array}\right\}
\left(\begin{array}{c}
e^{a}\\
f^{a}
\end{array}\right)=0, 
\end{equation}
with
\begin{eqnarray}
\mbox{\boldmath$e$} & = & \frac{\partial\n}{\partial\theta}= 
(\cos\theta\cos\varphi,\ \cos\theta\sin\varphi,\ -\sin\theta),\nonumber\\
\mbox{\boldmath$f$} & = & \frac{1}{\sin\theta}\frac{\partial\n}{\partial\varphi}=
(-\sin\varphi,\ \cos\varphi,\ 0).
\end{eqnarray}
In Appendix A, we give the details of the derivation of Eq. (2$\cdot$14).

\subsection{Existence of $\n(x)$}

In the previous subsections, we obtained the consistency condition 
(2$\cdot$11) and its integrability condition (2$\cdot$14).
Here we obtain a sufficient and very loose condition for the relation
(2$\cdot$14) to be satisfied.
Thus we realize that we are allowed to assume (1$\cdot$1) quite generally.

We first observe that the condition (2$\cdot$14) is satisfied
manifestly if the quantities
\begin{subequations}
\begin{equation}
H_{\mu\nu}^{a}\equiv\partial_\mu(\beta A_{\nu}^{a})-
\partial_\nu(\beta A_{\mu}^{a})+
\epsilon^{abc}A_{\mu}^{b}A_{\nu}^{c}(-2\beta\gamma)
\end{equation}
and
\begin{equation}
\hspace{3cm}G_{\mu\nu}^{a}\equiv\partial_\mu(\gamma A_{\nu}^{a})-
\partial_\nu(\gamma A_{\mu}^{a})+
\epsilon^{abc}A_{\mu}^{b}A_{\nu}^{c}(\beta^{2}-\gamma^{2})
\end{equation}
\end{subequations}
can be rewritten in the form,
\begin{subeqnarray}
H_{\mu\nu}^a\hspace{-3.5mm} & = & \hspace{-3.5mm}
n^{a}\xi_{\mu\nu}+e^{a}\zeta_{\mu\nu}+f^{a}\omega_{\mu\nu},\\
G_{\mu\nu}^{a}\hspace{-3.5mm} & = & \hspace{-3.5mm}
n^{a}\eta_{\mu\nu}+e^{a}\omega_{\mu\nu}-f^{a}\zeta_{\mu\nu}, 
\end{subeqnarray}
where $\xi_{\mu\nu}(x)$, $\eta_{\mu\nu}(x)$, $\zeta_{\mu\nu}(x)$ and 
$\omega_{\mu\nu}(x)$ 
are arbitrary anti-symmetric tensor fields.
Here we have made use of the relations $n^{a}(x)n^{a}(x)=e^{a}(x)e^{a}(x)=
f^{a}(x)f^{a}(x)=1$ 
and $n^{a}(x)e^{a}(x)=e^{a}(x)f^{a}(x)=f^{a}(x)n^{a}(x)=0$.
The most general form for $\xi_{\mu\nu}(x)$ appropriate in the present case 
is given by
\begin{eqnarray}
\xi_{\mu\nu} & = & (n^{a}\xi_{1}+e^{a}\xi_{2}+f^{a}\xi_{3})R_{\mu\nu}^{a}
+(n^{a}\xi'_{1}+e^{a}\xi'_{2}+f^{a}\xi'_{3})P_{\mu\nu}^{a}\nonumber\\
& { } & \hspace{1.5cm}+(n^a\xi''_1+e^a\xi''_2+f^a\xi''_3)Q_{\mu\nu}^a\nonumber\\
& \equiv &l_{\mu\nu}(\xi, \xi', \xi''),\nonumber\\
R_{\mu\nu}^a & = & \epsilon^{abc}A_\mu^b A_\nu^c,\nonumber\\
P_{\mu\nu}^a & = & \partial_\mu(\beta A_\nu^a)-\partial_\nu(\beta A_\mu^a),\nonumber\\
Q_{\mu\nu}^a & = & \partial_\mu(\gamma A_\nu^a)-\partial_\nu(\gamma A_\mu^a),\nonumber\\
\xi & = & (\xi_1, \xi_2, \xi_3), \hspace{3mm}  
\xi'=(\xi'_1, \xi'_2, \xi'_3),\hspace{3mm}
\xi''=(\xi''_1, \xi''_2, \xi''_3).
\end{eqnarray} 
Similarly, $\eta_{\mu\nu}(x)$, $\zeta_{\mu\nu}(x)$ and $\omega_{\mu\nu}(x)$ 
can be written as
\begin{equation} 
\eta_{\mu\nu}=l_{\mu\nu}(\eta, \eta', \eta''),\hspace{3mm} 
\zeta_{\mu\nu}=l_{\mu\nu}(\zeta, \zeta', \zeta''),\hspace{3mm} 
\omega_{\mu\nu}=l_{\mu\nu}(\omega, \omega', \omega'').
\end{equation}
Equating the r.h.s of (2$\cdot$16a) with that of (2$\cdot$17a), we have
\begin{subequations}
\begin{equation}
(-2\beta\gamma\delta^{ab}-U^{ab})R_{\mu\nu}^{b}+(\delta^{ab}-S^{ab})
P_{\mu\nu}^{b}-X^{ab}Q_{\mu\nu}^{b}=0.
\end{equation}
Similarly, (2$\cdot$16b) and (2$\cdot$17b) yield
\begin{equation}
\{(\beta^{2}-\gamma^{2})\delta^{ab}-V^{ab}\}R_{\mu\nu}^{b}-Y^{ab}
P_{\mu\nu}^{b}+(\delta^{ab}-T^{ab})Q_{\mu\nu}^{b}=0.
\end{equation}
\end{subequations}
In (2$\cdot$20), the functions $U^{ab}(x)$, $S^{ab}(x)$, $X^{ab}(x)$, 
$V^{ab}(x)$, $Y^{ab}(x)$ and $T^{ab}(x)$ are defined by
\begin{eqnarray}
S^{ab} & = & (n^a\ e^a\ f^a)
\left(\begin{array}{ccc}
\xi'_1 & \xi'_2 & \xi'_3\\
\zeta'_1 & \zeta'_2 & \zeta'_3\\
\omega'_1 & \omega'_2 & \omega'_3
\end{array}\right)
\left(\begin{array}{c}
n^b\\e^b\\f^b
\end{array}\right)
\equiv m^{ab}(\xi', \zeta', \omega'),\nonumber\\
\nonumber\\
T^{ab} & = & m^{ab}(\eta', \omega',-\zeta'),\hspace{3mm}
U^{ab}=m^{ab}(\xi, \zeta, \omega),\hspace{3mm}
V^{ab}=m^{ab}(\eta, \omega, -\zeta),\nonumber\\
\nonumber\\
X^{ab} & = & m^{ab}(\xi'', \zeta'', \omega''),\hspace{3mm}
Y^{ab}=m^{ab}(\eta'', \omega'', -\zeta'').
\end{eqnarray}
Defining the 3$\times$3 matrices $S(x)$, $T(x)$, $U(x)$, $V(x)$, $X(x)$ and 
$Y(x)$ and 3-vectors $R_{\mu\nu}(x)$, $P_{\mu\nu}(x)$ and $Q_{\mu\nu}(x)$ by
\begin{eqnarray}
S & = & \left(\begin{array}{ccc}
S^{11} & S^{12} & S^{13}\\
S^{21} & S^{22} & S^{23}\\
S^{31} & S^{32} & S^{33}
\end{array}\right),\hspace{3mm}\mbox{\rm etc}.,\nonumber\\
R_{\mu\nu} & = & \left(\begin{array}{c}
R_{\mu\nu}^1\\
R_{\mu\nu}^2\\
R_{\mu\nu}^3
\end{array}\right),\hspace{3mm}
\mbox{\rm etc}.,
\end{eqnarray} 
(2$\cdot$20a) and (2$\cdot$20b) become
\begin{subeqnarray}
& ( &\hspace{-3.5mm}-2\beta\gamma-U)R_{\mu\nu}+(1-S)P_{\mu\nu}-XQ_{\mu\nu}=0,\\ 
&\{ &\hspace{-3.5mm}(\beta^{2}-\gamma^{2})-V\}R_{\mu\nu}-YP_{\mu\nu}+(1-T)Q_{\mu\nu}=0.
\end{subeqnarray}
We find that the system (2$\cdot$23) or (2$\cdot$20) consists of 
thirty-six equations, since $a$ and $\mu\nu$ range over the values 
$a\ms=\ms1,\ 2,\ 3$ and 
$\mu\nu\ms=\ms01,\ 02,\ 03,\ 12,\ 23,\ 31$,
and that this system contains the thirty-six functions $\xi_i,\   
\eta_i,\ \zeta_i,\ \omega_i,\ \xi'_i,\ \eta'_i,\ \zeta'_i,\ \omega'_i,\ 
\xi''_i,\ \eta''_i,\ \zeta''_i$ and $\omega''_i \ (i\ms=\ms1,\ 2,\ 3)$.
It is expected that we can adjust these thirty-six functions so that the 
thirty-six equations in (2$\cdot$23) are satisfied.
We explicitly show in Appendix B that this is the case for general $R_{\mu\nu},
P_{\mu\nu}$ and $Q_{\mu\nu}$ such that
\begin{equation} 
\det W\neq0,
\end{equation}
where the 6$\times$6 matrix $W$ is defined by
\begin{eqnarray}
W=\left(\begin{array}{cclccc}
\vspace{2mm}
R_{01}^1 & R_{01}^2 & R_{01}^3 & P_{01}^1 & P_{01}^2 & P_{01}^3\\
R_{02}^1 & R_{02}^2 & R_{02}^3 & P_{02}^1 & \cdots\cdots \\
R_{03}^1 & R_{03}^2 & R_{03}^3 &  \cdots\cdots & & \vdots\\
R_{12}^1 & R_{12}^2 & \cdots\cdots & & \vdots\\
\vspace{2mm}
R_{23}^1 & \cdots\cdots & & \vdots & \cdots\cdots\cdots & P_{23}^3\\
R_{31}^1 & \cdots\cdots\cdots & &\cdots\cdots & P_{31}^2 & P_{31}^3 \\
\end{array}\right).\nonumber\\
\end{eqnarray}
Similar arguments lead us to sufficient conditions $\det W'(x)\neq0$ and
$\det W''(x)\neq0$, where the $6\times6$ matrices $W'(x)$ and $W''(x)$ are defined
by (2$\cdot$25) with $(R^a_{\mu\nu}(x),\ P^a_{\mu\nu}(x))$ replaced by
$(P^a_{\mu\nu}(x),\ Q^a_{\mu\nu}(x))$ and $(R^a_{\mu\nu}(x),\ Q^a_{\mu\nu}(x))$,
respectively. These conditions are quite loose.
Roughly speaking, gauge configrations that satisfy none of $\det W(x)\neq0$,
$\det W'(x)\neq0$, $\det W''(x)\neq0$ constitute, at most, a set of zero
measure in the space of gauge configrations. Since these conditions are sufficient
but not necessary for the existence of $\n(x)$, the fact might be that the field
$\n(x)$ exists even if $\det W(x)$, $\det W'(x)$ and $\det W''(x)$ all vanish. 
For example, with $\rho(x)=0$, $\sigma(x)=-1$ and $C_\mu(x)=0$,
$\n(x)$ can be arbitrary for the gauge configuration 
$\A_{\mu}(x)=0$ which implies $\det W(x)=\det W'(x)=\det W''(x)=0$.

Now that we have shown that the integrability condition is satisfied quite 
generally, we can determine $\n(x)$ from (2$\cdot$11) in principle.
Compared with the original form of the consistency conditions 
(i.e., (2$\cdot$3) and (2$\cdot$4)), Eq. (2$\cdot$11), or more explicitly,
\begin{equation}
\left(\begin{array}{c}
\partial_\mu\theta\\
\sin\theta\hspace{1mm}\partial_\mu\varphi
\end{array}\right)
=\left(\begin{array}{cc}
\beta & \gamma\\
-\gamma & \beta
\end{array}\right)
\left(\begin{array}{c}
A^1_\mu\cos\theta\cos\varphi+A^2_\mu\cos\theta\sin\varphi-A^3_\mu\sin\theta\\
-A^1_\mu\sin\varphi+A^2_\mu\cos\varphi
\end{array}\right)
\end{equation}
seems much easier to deal with. For a given gauge configuration $\A_\mu(x)$
and for a given pair of arbitrary functions $\beta(x)$ and $\gamma(x)$,
the field $\n(x)$ is obtained by solving the above partial differential equation.
We thus answer the questions (1) and (2) posed in \S1.  
We see that both of the fields $\n(x)$ and $\A_{\mu}(x)$ given by (1$\cdot$1)
can be regarded as functionals of $\beta(x)$ and $\gamma(x)$ and satisfy 
(2$\cdot$12).
 On the other hand, gauge transformations caused by a unitary matrix $U
[\alpha]\equiv$ exp$[i\alpha(x)n^{a}(x)T^a]$, where $T^a$ is a representation 
of the generator of $SU(2)$, leaves $\n(x)$ invariant but gives rise to 
transformations of $C_{\mu}(x)$, $\rho(x)$ and $\sigma(x)$. \cite{rf:1} \
 We then suppose that the functions $\alpha(x), \ \beta(x)$ and $\gamma(x)$ 
play the role of three gauge functions of the $SU(2)$ gauge transformations.

\section{The case of $SU(N)$}

As stated in \S1, we consider Periwal's ansatz for the $SU(N)$ gauge field
and its consistency conditions. We also discuss that his ansatz causes some
conditions on the gauge field.

\subsection{Periwal's ansatz}

Periwal's extension of the Faddeev-Niemi Ansatz 
(1$\cdot$1) to the case of $SU(N)$ is given by
\begin{equation}
A_\mu=C_{\mu,I} H_I+\phi_I{\mathcal D}_\mu H_I
+i\beta_{IJ}[H_I,{\mathcal D}_\mu H_J]+\mbox{(nonlinear terms)},
\end{equation}
where $A_\mu(x)$ is a gauge-fixed massless $SU(N)$ gauge field,
$C_{\mu,I}(x)$ is a gauge-fixed massless Abelian gauge field, and 
$\phi_I(x)$ and $\beta_{IJ}(x)\ms=\ms\beta_{JI}(x)$ 
are scalar fields $(I,J\ms=\ms1\ ,2,\ \cdots,\ N-1)$.
In (3$\cdot$1) summation over $1,\ 2,\ \cdots,\ N-1$ 
is implied by the repeated indices $I$ and $J$.
The field $\n(x)$ for the $SU(2)$ case has been replaced by $N-1$ commuting
traceless Hermitian $N\times N$ matrices $H_I(x)$ $(I\ms=\ms1,\ 2,\ \cdots,\ N-1)$.
The covariant derivertive ${\mathcal D}_\mu$ in (3$\cdot$1) is defined by
\begin{eqnarray}
{\mathcal D}_\mu H_I & \equiv & \partial_\mu H_I+\omega_{\mu,IJ}H_J,\nonumber\\
\omega_{\mu,IJ} & \equiv & \mbox{\rm tr}(H_I\partial_\mu H_J).
\end{eqnarray} 
Periwal assumed that there appears no new field other than $C_{\mu,I}(x)$, 
$H_I(x)$, $\phi_I(x)$
and $\beta_{IJ}(x)$ in the (nonlinear terms) of (3$\cdot$1).
To keep simple transformation laws of $C_{\mu,I}(x)$, $\phi_I(x)$ and $\beta_{IJ}(x)$
under a class of gauge transformations, he proposed that the complete form 
including (nonlinear terms) is given by 
\begin{equation}
A_\mu={\rm Ad}_\phi(C_{\mu,I}H_I+i\beta_{IJ}[H_I,{\mathcal D}_\mu H_J]+E_\mu)-E_\mu,
\end{equation}
where $E_\mu(x)$ is defined as a matrix satisfying 
$-i[H_I,E_\mu]\equiv{\mathcal D}_\mu H_I$, and Ad$_\phi$ represents
Ad$_\phi(X)\equiv e^{-i\phi_IH_I}Xe^{i\phi_IH_I}$.

The degrees of freedom contained in $\{H_I(x)|I\ms=\ms1,\ 2,\ \cdots,\ N-1\}$ can be
seen in the following way. First we fix an $N$-dimensional representation 
of $SU(N)$. Its $x$-independent $N^2-1$ generators are divided 
into $h_I$ $(I\ms=\ms1,\ 2,\ \cdots,\ N-1)$
and $t_a$ $(a\ms=\ms1,2,\cdots,N^2-N)$, 
where $h_I$ is one of $N-1$ generators 
of the Cartan subalgebra of $SU(N)$, and $t_a$ is one of the residual $N^2-N$
generators. It is known that the most general expression for $H_I(x)$ 
takes the form 
\begin{equation}
H_I=U\alpha_{IJ}h_JU^\dagger,
\end{equation}
where $\alpha_{IJ}(x)$ $(I,J\ms=\ms1,\ 2,\ \cdots,\ N-1)$ are real functions, 
and the matrix $\alpha(x)\ms\equiv\ms(\alpha_{IJ}(x))$ is assumed to belong to
$SO(N-1)$. The matrix $U(x)$ in (3$\cdot$4) is given by 
\begin{equation}
U=e^{i\gamma_1t_1}e^{i\gamma_2t_2}\cdots e^{i\gamma_{N^2-N}t_{N^2-N}}
\end{equation} 
with $\gamma_a(x)$ $(a\ms=\ms1,\ 2,\ \cdots,\ N^2-N)$ being real functions.
Noting that $\dim(SO(N-1))=(N-1)(N-2)/2$, it is now clear 
that the degrees of freedom of $\{H_I(x)|I\ms=\ms1,\ 2,\ \cdots,\ N-1\}$ is equal to
$N^2-N+(N-1)(N-2)/2$. As explained in \S1, the sum of the number of degrees of 
freedom of $H_I(x)$, $C_{\mu,I}(x)$, $\phi_I(x)$ and $\beta_{IJ}(x)$ is
equal to $2(N^2-1)$. This is equal to the number of degrees of freedom of a 
gauge-fixed massless $SU(N)$ gauge field. 
We shall see in the following, however, that the $(N-1)(N-2)/2$ degrees of 
freedom contained in $\alpha_{IJ}(x)$ are not independent of the other fields
in (3$\cdot$3).
  
\subsection{Consistency conditions implied by Periwal's ansatz}

We now seek the consistency conditions of the ansatz (3$\cdot$3), 
which are analogous to (2$\cdot$3) and (2$\cdot$4) in the $SU(2)$ case. 
We first note that,
from the definitions (3$\cdot$2) and the expression (3$\cdot$4), we obtain 
\begin{eqnarray}
\omega_{\mu,IJ} & = & -\alpha_{JK}\partial_\mu\alpha_{IK},\\
{\mathcal D}_\mu H_I & = & [(\partial_\mu U)U^\dagger,H_I],
\end{eqnarray}
and hence $E_\mu$ in (3$\cdot$3) is given by 
\begin{equation}
E_\mu=iU(\partial_\mu U^\dagger).
\end{equation}
It is noteworthy that the derivative $\partial_\mu\alpha_{IJ}$ disappears 
in ${\mathcal D}_\mu H_I$. It is straightforward to obtain the orthogonality
relations
\begin{subeqnarray}
& \mbox{\rm tr} & \hspace{-4mm}\{({\mathcal D}_\mu H_I)
[H_J,{\mathcal D}_\mu H_K]\}=0,\\
& \mbox{\rm tr} & \hspace{-4mm}\{H_I[H_J,{\mathcal D}_\mu H_K]\}=0,\\
& \mbox{\rm tr} & \hspace{-4mm}\{H_I({\mathcal D}_\mu H_J)\}=0,
\end{subeqnarray}
the last of which is the requirement adopted by Periwal to fix $\omega_{\mu,IJ}$,
as in (3$\cdot$2). From (3$\cdot$3), we easily obtain 
\begin{equation}
C_{\mu,I}=\mbox{\rm tr}(H_IA_\mu).
\end{equation}
Defining $(B_\mu)_{IJ,LM}$ by 
\begin{equation}
(B_\mu)_{IJ,LM}=\mbox{\rm tr}([H_I,{\mathcal D}_\mu H_J]
[H_L,{\mathcal D}_\mu H_M])
\end{equation}
and regarding it as the $(IJ,LM)$ element of the $(N-1)^2\times(N-1)^2$
matrix $B_\mu$, we obtain the following formula for $i\beta_{IJ}$ 
from (3$\cdot$3) and (3$\cdot$9):
\begin{equation}
i\beta_{IJ}=\mbox{\rm tr}([H_L,{\mathcal D}_\mu H_M]
\{\mbox{\rm Ad}_{-\phi}(A_\mu+E_\mu)-E_\mu\})(B^{-1}_\mu)_{LM,IJ}
\equiv\gamma_{IJ,\mu}
\end{equation}
Similarly, we have the relation which relates $\phi_I$ to $A_\mu$ and $H$:
\begin{equation}
\mbox{\rm tr}([E_\mu,H_I]\mbox{\rm Ad}_{-\phi}(A_\mu+E_\mu))=0,
\end{equation}
or more explicitly
\begin{eqnarray}
& \phi & _I=\mbox{\rm tr}(A_\mu{\mathcal D}_\mu H_J)(K^{-1}_\mu)_{IJ}
+\cdots\cdots\equiv\psi_{I,\mu},\\
( & K & _\mu)_{IJ}=\mbox{\rm tr}\{({\mathcal D}_\mu H_I)({\mathcal D}_\mu H_J)\}.
\end{eqnarray}
We note that the summation over the repeated index $\mu$ is not implied in
(3$\cdot$11)-(3$\cdot$15) but that $\mu$ should be fixed as $\mu\ms=\ms0$, 1, 2 or 3.
For the Periwal ansatz to be consistent, $\gamma_{IJ,\mu}$ and $\psi_{I,\mu}$
defined in (3$\cdot$12) and (3$\cdot$14), respectively, must satisfy
\begin{eqnarray}
\gamma_{IJ,0} & = & \gamma_{IJ,1}=\gamma_{IJ,2}=\gamma_{IJ,3},\hspace{3mm}I\geq J,\\
\psi_{I,0} & = & \psi_{I,1}=\psi_{I,2}=\psi_{I,3}.
\end{eqnarray}
Thus we have obtained $3N(N-1)/2+3(N-1)=3(N-1)(N+2)/2$ consistency conditions.

We next investigate if all the fields $\{\alpha_{IJ}(x)\}$, $\{\gamma_a(x)\}$,
$\{C_{\mu,I}(x)\}$, $\{\phi_I(x)\}$ and $\{\beta_{IJ}(x)\}$
can be regarded as independent. 
After the conditions (3$\cdot$16) and (3$\cdot$17) are imposed, the ansatz (3$\cdot$3)
is insured if it is satisfied for one of $\mu=0$, 1, 2, 3.
Thus we consider an equation with the suffix $\mu$ in (3$\cdot$3) omitted:
\begin{equation}
A=\mbox{\rm Ad}_\phi(C_IH_I+i\beta_{IJ}[H_I,{\mathcal D}H_J]+E)-E.
\end{equation}
From the structure of the r.h.s. of (3$\cdot$18), and noting the comment below
(3$\cdot$8), we observe that there exists a hidden $SO(N-1)$ symmetry
in (3$\cdot$18). Expressing $H_I(x)$ in the form of (3$\cdot$4), $A$ is
written as $A(\{C_I\}$, $\{\phi_I\}$, $\{\beta_{IJ}\}$, $U$, $\{\alpha_{IJ}\})$, 
which does not contain derivatives of $C_I$, $\phi_I$, $\beta_{IJ}$ 
and $\alpha_{IJ}$.Then we have 
\begin{eqnarray}
A(\{\xi_{IJ}C_J\},\ \{\xi_{IJ}\phi_J\},\ \{\xi_{IL}\xi_{JM}\beta_{LM}\},
\ U,\ \{\xi_{IK}\alpha_{KJ}\})\nonumber\\
=A(\{C_I\},\ \{\phi_I\},\ \{\beta_{IJ}\},\ U,\ \{\alpha_{IJ}\}),
\end{eqnarray}
where $\xi_{IJ}$ is the $IJ$-element of an arbitrary matrix $\xi$ belonging to
$SO(N-1)$. By setting $\xi=e^{i\theta_\alpha J_\alpha}$, with $\theta_\alpha$
and $J_\alpha$ being an infinitesimal group parameter and an $N-1$ dimensional
representation of the generator of $SO(N-1)$, respectively, we obtain 
the $(N-1)(N-2)/2$ conditions
\begin{eqnarray}
\frac{\partial A}{\partial C_I}(J_\alpha)_{IJ}C_J
+\frac{\partial A}{\partial\phi_I}(J_\alpha)_{IJ}\phi_J
+\frac{\partial A}{\partial\alpha_{IJ}}(J_\alpha)_{IK}\alpha_{KJ}\nonumber\\
+\frac{\partial A}{\partial\beta_{IJ}}
\{(J_\alpha)_{IK}\beta_{KJ}+(J_\alpha)_{JK}\beta_{IK}\}=0,\nonumber\\
\alpha=1,\ 2,\ \cdots,\ \frac{(N-1)(N-2)}{2}.
\end{eqnarray}
Setting $\xi(x)\equiv(\xi_{IJ}(x))=\alpha(x)^{-1}$ in (3$\cdot$19),
we see that the genuine independent degrees of freedom in the field $A(x)$
are $\{C_I(x)\}$, $\{\phi_I(x)\}$, $\{\beta_{IJ}(x)\}$ and $U(x)$ and that
the shortage of degrees of freedom gives rise to the conditions (3$\cdot$20).

\section{Summary}

In this paper, we have investigated some aspects of the Faddeev-Niemi ansatz 
proposed recently for a gauge-fixed massless $SU(2)$ gauge field $\A_\mu(x)$. 
We have discussed that some consistency conditions must be imposed on 
$\A_\mu(x)$ and $\n(x)$ and that these conditions can be used to fix the 
field $\n(x)$. We have written the consistency conditions as a system of 
differential equations for $\n(x)$ and found that the integrability conditions 
of the system is satisfied for gauge configurations obeying (2$\cdot$24).
The condition (2$\cdot$24) is rather loose. Since it is a sufficient 
but not a necessary condition, it may be the case that the ansatz can be used generally.
It has been observed also that the solution $\n(x)$ of the consistency 
condition (2$\cdot$26) contains two arbitrary functions $\beta(x)$ and $\gamma(x)$ 
which are related to $\A_\mu(x)$ by (2$\cdot$8).   
These functions might play the role of two of the three gauge functions 
of the $SU(2)$ gauge group.

We have sought the consistency conditions for the Periwal ansatz for gauge-fixed
massless $SU(N)$ gauge field and found some conditions.
We have seen that some of the fields introduced in his ansatz cannot be regarded as 
independent.
\section*{Acknowledgements}

We are grateful to Shinji Hamamoto, Takeshi Kurimoto and 
Makoto Nakamura for valuable comments.

\appendix

\section{Derivation of (2$\cdot$14)}

Differentiating (2$\cdot$11), we have 
\begin{eqnarray}
\partial_\nu 
\left\{\begin{array}{c}
L\left(\begin{array}{c}
G_\mu\\
H_\mu
\end{array}\right)
\end{array}\right\}
& = & \partial_\nu\left(\begin{array}{c}
                  \partial_\mu\theta\\
                  \sin\theta\partial_\mu\varphi
                  \end{array}\right)\nonumber\\
& = & \left(\begin{array}{c}
                  \partial_\nu(\partial_\mu\theta)\\
                  \sin\theta\partial_\nu(\partial_\mu\varphi)
                  \end{array}\right) 
     +\left(\begin{array}{c}
      0\\
      \cos\theta(\partial_\nu\theta)(\partial_\mu\varphi)
      \end{array}\right).
\end{eqnarray}
Making use of (2$\cdot$11), the second term on the r.h.s. of 
(A$\cdot$1) becomes 
\begin{eqnarray}
\left(\begin{array}{c}
0\\
\cos\theta(\partial_\nu\theta)(\partial_\mu\varphi)
\end{array}\right)\hspace{7cm}\nonumber\\
=\frac{\cos\theta}{\sin\theta}
\left(\begin{array}{c}
0\\
\beta^2G_\nu H_\mu-\gamma^2H_\nu G_\mu+\beta\gamma(H_\mu H_\nu-G_\mu G_\nu)
\end{array}\right).
\end{eqnarray}
Requiring (2$\cdot$13), we obtain
\begin{eqnarray}
\partial_\nu
\left(\begin{array}{c}
\partial_\mu\theta\\
\sin\theta\partial_\mu\varphi
\end{array}\right)
& - & \partial_\mu
 \left(\begin{array}{c}
 \partial_\nu\theta\\
 \sin\theta\partial_\nu\varphi
 \end{array}\right)\nonumber\\
& = & \frac{\cos\theta}{\sin^2\theta}(\beta^2+\gamma^2)A^a_\nu A^b_\mu
  \biggl(\frac{\partial n^a}{\partial\theta}\frac{\partial n^b}{\partial\varphi}
  -\frac{\partial n^b}{\partial\theta}\frac{\partial n^a}{\partial\varphi}\biggr).
\end{eqnarray}
On the other hand, the l.h.s. of (A$\cdot$1) is given by
\begin{equation}
\partial_\nu
\left\{\begin{array}{c}
L\left(\begin{array}{c}
G_\mu\\
H_\mu
\end{array}\right)
\end{array}\right\}
=\{\partial_\nu(LA^a_\mu)\}
  \left(\begin{array}{c}
  \frac{\partial n^a}{\partial\theta}\\
  \frac{1}{\sin\theta}\frac{\partial n^a}{\partial\varphi}
  \end{array}\right)+D_{\nu\mu}
\end{equation}
with
\begin{equation}
D_{\nu\mu}=LA^a_\mu
 \left(\begin{array}{cc}
 \frac{\partial^2n^a}{\partial\theta^2}
 &\frac{\partial^2n^a}{\partial\varphi\partial\theta}\\
 \frac{\partial}{\partial\theta}
 \bigl(\frac{1}{\sin\theta}\frac{\partial n^a}{\partial\varphi}\bigl)
 &\frac{\partial}{\partial\varphi}
 \bigl(\frac{1}{\sin\theta}\frac{\partial n^a}{\partial\varphi}\bigl)
 \end{array}\right)
  \left(\begin{array}{c}
  \partial_\nu\theta\\
  \partial_\nu\varphi
  \end{array}\right).
\end{equation}
With the help of (2$\cdot$11), the r.h.s. of (A$\cdot$5) can be expressed 
in terms of $A^a_\mu$, $L$, $\theta$ and $\varphi$.
Although the expressions for $D_{\nu\mu}$ and $D_{\nu\mu}-D_{\mu\nu}$
are complicated, we finally obtain
\begin{eqnarray}
\left\{\begin{array}{c}
\partial_\nu\left(\begin{array}{c}
                   \partial_\mu\theta\\
                   \sin\theta\partial_\mu\varphi
                   \end{array}\right)
-\partial_\mu\left(\begin{array}{c}
                   \partial_\nu\theta\\
                   \sin\theta\partial_\nu\varphi
                   \end{array}\right)
\end{array}\right\}
-(D_{\nu\mu}-D_{\mu\nu})\nonumber\\
=-\epsilon^{abc}A^a_\mu A^b_\nu
\left(\begin{array}{cc}
2\beta\gamma & -(\beta^2-\gamma^2)\\
\beta^2-\gamma^2 & 2\beta\gamma
\end{array}\right)
\left(\begin{array}{c}
e^c\\
f^c
\end{array}\right),
\end{eqnarray}
where $e^c$ and $f^c$ are components of the vectors 
$\mib{e}$ and $\mib{f}$ in (2$\cdot$15). 
It is now easy to obtain (2$\cdot$14) from the above relations.

\section{Existence of thirty-six functions insuring (2$\cdot$24)}

We here show that we can indeed adjust the functions 
$\xi_i$, $\eta_i$, $\zeta_i$, $\omega_i$, $\xi'_i$, $\eta'_i$, $\zeta'_i$, $\omega'_i$, 
$\xi''_i$, $\eta''_i$, $\zeta''_i$ and $\omega''_i$ $(i\ms=\ms1,\ 2,\ 3)$ so that 
(2$\cdot$23a) and (2$\cdot$23b) are satisfied. We first note that the matrices
$T$, $V$ and $Y$ can be written as
\begin{equation}
T=JS+I',\hspace{3mm}V=JU+I,\hspace{3mm}Y=JX+I'',
\end{equation}
where the matrix elements of the matrices $J$, $I$, $I'$ and $I''$ 
are given by
\begin{eqnarray}
J^{ab} & = & n^an^b+e^af^b-f^ae^b,\hspace{3mm}
I^{ab}=m^{ab}(\eta-\xi,0,0),\nonumber\\
I'^{ab} & = & m^{ab}(\eta'-\xi',0,0),\hspace{3mm}
I''^{ab}=m^{ab}(\eta''-\xi'',0,0).
\end{eqnarray}
With the help of (B$\cdot$1), we see that (2$\cdot$23a) is equivalent to
(2$\cdot$23b) if we can write 
\begin{subeqnarray}
-2\beta\gamma-U\hspace{-3mm} & = & \hspace{-3mm}D\{(\beta^2-\gamma^2)-(JU+I)\},\\
1-S\hspace{-3mm} & = & \hspace{-3mm}-D(JX+I''),\\
-X\hspace{-3mm} & = & \hspace{-3mm}D(1-JS-I')
\end{subeqnarray}
and the matrix $D$ is invertible. We regard (B$\cdot$3a) as the definition
of $D$. It is clear that an appropriate choice of $U$ and $I$ yields a 
well-defined and invertible $D$. We obtain $X=D(JS+I'-1)$ from (B$\cdot$3c) and
\begin{equation}
I''=D^{-1}(S-1)-JX
\end{equation}
from (B$\cdot$3b). These relations fix $\xi''_i$, $\eta''_i$, $\zeta''_i$
and $\omega''_i$ $(i\ms=\ms1,\ 2,\ 3)$ in terms of 
$\Xi\equiv\{\xi_i$, $\eta_i$, $\zeta_i$, $\omega_i$, 
$\xi'_i$, $\eta'_i$, $\zeta'_i$, $\omega'_i
|i\ms=\ms1,\ 2,\ 3\}$.
Since the six zeroes in $I''$ of (B$\cdot$4) impose six restrictions on the
members of $\Xi$, only eighteen of them are independent. We are then left 
with eighteen equations for the above eighteen functions:
\begin{equation}
KR_{\mu\nu}+LP_{\mu\nu}=Q_{\mu\nu},
\end{equation}
with $K$ and $L$ given by 
\begin{eqnarray}
K & = & X^{-1}(-2\beta\gamma-U),\nonumber\\
L & = & X^{-1}(1-S).
\end{eqnarray}
If we define column vectors $x$ and $y$ by
\begin{eqnarray}
\hspace{-10mm}x & = & ^t(K^{11},\ K^{12},\ K^{13},\ L^{11},\ L^{12},\ L^{13},\ 
           K^{21},\ K^{22},\ \cdots,\ L^{32},\ L^{33}),\nonumber\\
\hspace{-10mm}y & = & ^t(Q^1_{01},\ Q^1_{02},\ Q^1_{03},\ Q^1_{12},\ Q^1_{23},\ Q^1_{31},\ 
           Q^2_{01},\ \cdots,\ Q^2_{31},\ Q^3_{01},\ \cdots,Q^3_{31}),
\end{eqnarray}
Eq. (B$\cdot$5) becomes
\begin{equation}
Zx=y,
\end{equation}
where the $18\times18$ matrix $Z$ is given by 
\begin{equation}
Z=\left(\begin{array}{ccc}
W&0&0\\
0&W&0\\
0&0&W
\end{array}\right).
\end{equation}
Here $W$ is the $6\times6$ matrix defined by (2$\cdot$25).
In the case that $\det Z=(\det W)^3\neq0$, we have 
\begin{equation} 
x=Z^{-1}y.
\end{equation}
The r.h.s. of (B$\cdot$10) is fixed by $R^a_{\mu\nu}$, $P^a_{\mu\nu}$, 
and $Q^a_{\mu\nu}$ and hence by $\A_\mu(x)$, $\beta(x)$ and $\gamma(x)$,
while the l.h.s. is fixed by the functions belonging to $\Xi$.
In other words, if we fix the functions $\xi''_i$, $\eta''_i$, $\zeta''_i$
and $\omega''_i$ $(i\ms=\ms1,2,3)$ by (B$\cdot$3) and $\Xi$ by (B$\cdot$4) and
(B$\cdot$10), both (2$\cdot$20a) and (2$\cdot$20b) are satisfied.

\end{document}